\documentclass[runningheads]{svmult}

\usepackage{makeidx}   
\usepackage{graphicx}  
\usepackage{subeqnar}  
\usepackage{multicol}  
\usepackage{physprbb}  
\makeindex             

\begin{document}
\title*{Evidence in Favour of IMF Variations}
\toctitle{Evidence in Favour of IMF Variations}
%
%
\titlerunning{Evidence in Favour of IMF Variations}
%
\author{Frank Eisenhauer}
\authorrunning{Frank Eisenhauer}
%
%
\institute{Max-Planck-Institut f\"ur extraterrestrische Physik, \\
Giessenbachstrasse, 85741 Garching, Germany}

\maketitle              

\begin{abstract}
The stellar initial mass function (IMF) determines the relative number
of stars born at a given mass. Despite the tremendous effort to
establish a universal IMF, the astronomical literature offers a wealth
of diverse evidence for IMF variations. This review was prepared for a
controversial debate at the conference ``Starbursts -- Near and Far''
at Ringberg Castle, 2000, and gives a one-sided portrayal in favour of
IMF variations. I will summarise the empirical evidence that the IMF
varies with time, with environment, and for all stellar masses. While
I see no obvious systematic trend in most regions of our Galaxy, there
is at least an indication that the IMF is biased towards more massive
stars in the early universe and in starbursts.
\end{abstract}

\section{Introduction}

The masses of stars span at least 3 orders of magnitude, from
approximately 100 solar masses for the highest mass stars to the
hydrogen burning limit of 0.1 solar masses. The relative numbers of
stars born at a given mass is described by the initial mass function
(IMF). This distribution is extremely important for many fields in
astronomy, from the theory of star formation to the interpretation of
integrated properties of galaxies at the highest redshift. Although
many investigators have measured the IMF in a variety of star forming
regions, we have not yet been able to come up with a final conclusion
about the universality of the IMF. Specifically, Scalo's article
\cite{scalo98} for the recent conference ``The Stellar Initial Mass
Function'' \cite{gilmore98} has revived the discussion, and many
scientists --- including the author --- have not decided on their
final opinion. Nevertheless the conference organisers have chosen to
split the review on the universality of the IMF to keep the discussion
lively and controversial. \emph{This contribution focuses strictly on
the empirical evidence in favour of IMF variations.} I will not try to
balance any arguments, but leave the critical comparison with the
evidence in favour of a universal IMF \cite{gilmore01} to the reader.

The postulation of a universal IMF --- in its strongest form ---
states that the IMF is and has always been the same in all regions of
star formation everywhere; the frequencies of initial stellar masses
in any unbiased sample are always drawn from the same statistical
distribution. I will present evidence against this postulation with
respect to ``is and has always been'' as well as to ``everywhere'',
and I will also argue against this postulation being valid even in
star forming regions of our own Galaxy.

\begin{figure}[ht]
\begin{center}
\includegraphics[width=\textwidth]{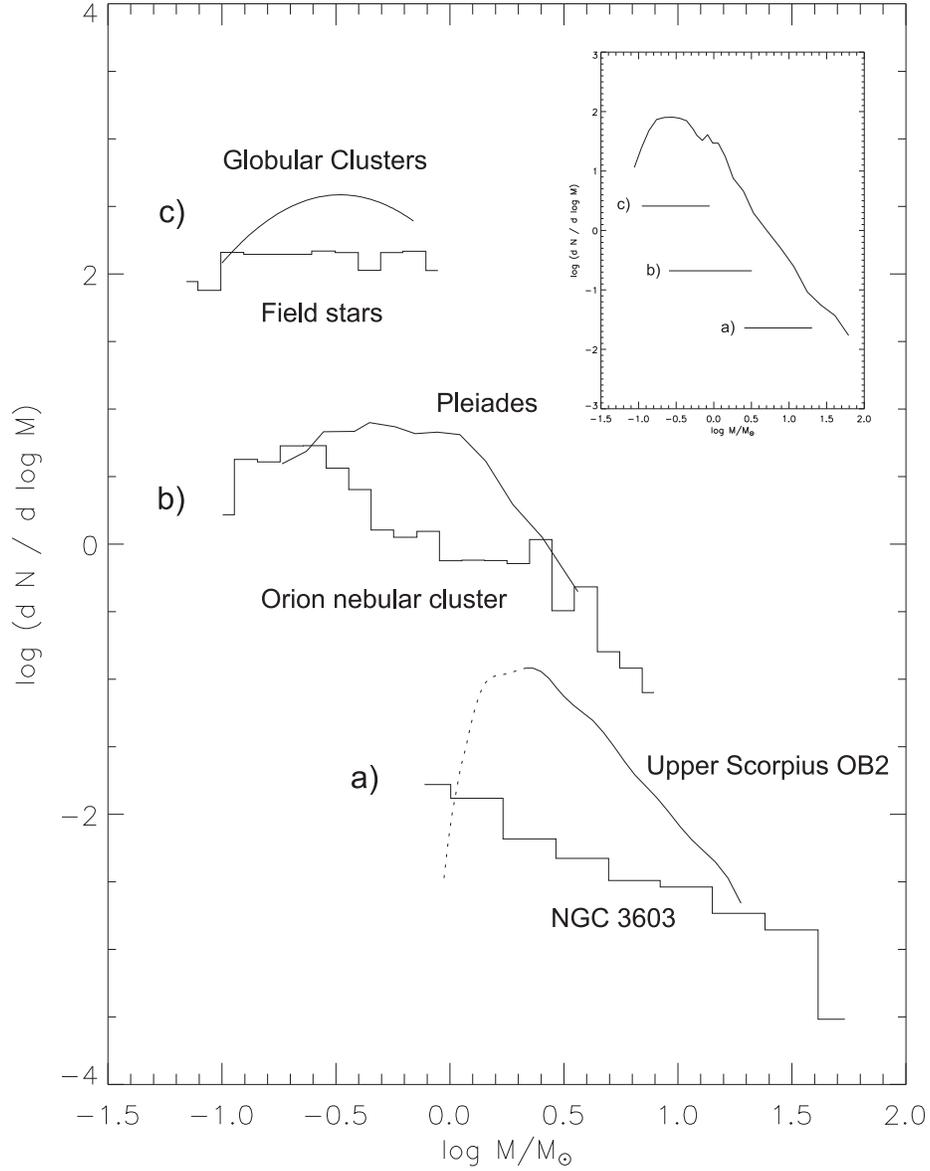}
\end{center}
\caption[]{Evidence for IMF variations: This figure shows a comparison
of the IMF of several selected Galactic regions.  Variations of the
IMF are evident for all stellar masses: a) The exponent of the IMF
differs by more than one in the high mass star forming regions
NGC~3603 \cite{eisenhauer98} and Upper Scorpius OB2 \cite{brown98}. b)
Compared to the Pleiades \cite{meusinger96}, the IMF of the Orion
nebular cluster \cite{hillenbrand97} is deficit in stars with masses
between 0.5 -- 1.25 $M_\odot$. c) The low-mass IMF in globular
clusters is fitted best by a log-normal distribution \cite{paresce00},
but the field stars follow a power law in the same mass interval
\cite{reid99}. The small inset shows the field star IMF of Scalo
\cite{scalo86} for comparison.  }
\label{figure}
\end{figure}

Although challenging the concept of a universal IMF, I refer to a
reference IMF for comparison. This IMF (small inset in figure
\ref{figure}) is the field star IMF defined by Scalo
\cite{scalo86} in 1986, because its qualitative behaviour seems to be
typical in many respects: (1) For high stellar masses the IMF can be
described by a power law $\frac{d\ N}{d\ log\ M} \propto
M^{\Gamma}$. The exponent is referred to as slope of the IMF. In the
notation of this article the value for the Salpeter \cite{salpeter55}
IMF is -1.35.  This power law results from scale free star formation
processes, for example dominated by turbulent pressure. (2) The IMF is
flat --- the slope of the power law approximately 0 --- for masses
around 0.5 $M_\odot$, indicating a characteristic stellar mass, for
example the thermal Jeans mass. (3) The IMF declines for low stellar
masses and may be described by a log-normal distribution, indicating a
large number of independent parameters in the star formation
process. However, I will present evidence that the IMF differs in
various regions from this reference IMF in shape (e.g. log-normal
versus power law and turnover versus continuous), in slope (e.g. bias
towards high or low masses) and in characteristic mass.

\section{The Early Universe}

If the IMF depends on any environmental conditions, one would expect
the largest deviations for the most extreme environments. The early
Universe is such an environment. There are two very compelling
indications that the IMF was different in the early Universe: First,
no metal-free stars and only few very-low-metallicity stars have ever
been found \cite{beers00,chiosi00}, although low-mass stars, which
have formed in the metal-poor early Universe, should have survived the
last 15 billion years. Only $\approx 500$ stars are known with [Fe/H]
$\approx \frac{1}{300}$ [Fe/H]$_\odot$ --- the lowest metallicity
observed in Galactic globular clusters ---, and only $\approx 100$
stars with [Fe/H] $\approx \frac{1}{1000}$ [Fe/H]$_\odot$. No single
star with primordial metallicity [Fe/H] $\approx \frac{1}{1000000}$
[Fe/H]$_\odot$ has ever been observed. Second, the solar neighbourhood
is deficient in metal-poor stars \cite{martinelli00}. This discrepancy
with the field star IMF is known as the ``classical G-dwarf
problem''. Because the lifetime of these stars is longer than the age
of the Universe, we would expect to find many more such stars, if
low-mass stars have formed with the same frequency in the early days
of our Galaxy as they do now. As outlined by Larson \cite{larson98},
this evidence finds its natural explanation in a time dependent IMF
with a bias towards massive stars in the early Universe. Such a
varying IMF is favoured for other reasons as well. For example, the
gas of galaxy clusters is very hot and contains a large mass of heavy
elements, which can not be explained from the proportionally few
massive stars of the present day IMF. Also the observed strong
evolution of the cosmic luminosity density with redshift is easier to
explain with a top-heavy IMF. And microlensing experiments 
indicate that a
significant fraction of dark matter is in the form of stellar
remnants, which can only be produced from massive progenitors. See
Larson's article \cite{larson98} for a more detailed explanation and
references.

\clearpage

\section{Starbursts and Extragalactic Star Forming Regions}

\subsection{Starburst Galaxies}

Starburst galaxies are another extreme environment for star formation.
Their star formation rate is several magnitudes larger than in our own
Galaxy. So if the thermal Jeans mass is of any importance for a
characteristic stellar mass, the heating from the unusually strong star
formation should give preference to more massive stars.

The starburst galaxy M~82 has been the prime example for a top heavy
IMF for 20 years \cite{rieke80}. As the closest starburst galaxy with
a distance of only 3.2 Mpc, no other starburst galaxy has been
observed in such detail. The appearance of this galaxy is dominated by
its nuclear starburst, which has probably been induced by a close
encounter with M~81 about 100 Myr ago. The basic properties of the
nuclear starburst are \cite{mcleod93}: Bolometric luminosity $> 5.0
\times 10^{10}$ $L_\odot$, absolute K magnitude $< -22.5$, ionising
radiation $L_{Lyc} > 10^{54}$ $photons/s$, total mass in the starburst
$< 2.5 \times 10^8$ $M_\odot$, CO index $> 0.21$ and supernova rate
$\approx 0.1$ $yr^{-1}$.  Rieke et al. \cite{rieke80,rieke93} have
pioneered the detailed modelling of this starburst galaxy, and have
come repeatedly to the conclusion that stars with masses below a few
$M_\odot$ form much less often in M~82 than in the solar
neighbourhood. The argument for a top heavy IMF is basically that the
infrared brightness --- tracing the bolometric luminosity --- is too
high for the (dynamically measured) stellar mass, if one assumes a IMF
similar to the field star IMF of Scalo \cite{scalo86}. The ionising
radiation, the CO index and the supernova rate provide the necessary
constraints for the age of the stellar population.

Although recent modelling \cite{satyapal97,foerster98} with a more
complex spatial structure and temporal evolution can not exclude a
Salpeter power-law IMF extending down to subsolar masses, the
starburst galaxy M 82 remains one of the most cited examples for IMF
variations.

\subsection{Extragalactic Super Star Clusters}

The same technique can also be applied to extragalactic super star
clusters. Spectral diagnostics, such as hydrogen emission lines and
stellar absorption features, enable us to date the starburst.
Dynamical mass estimates together with infrared observations provide a
reliable mass-to-light ratio, and these together constrain the shape
of the IMF.  Compared with starburst galaxies, individual super star
clusters offer the big advantage of a simple spatial (only one
cluster) and temporal (only one burst) structure. Nevertheless these
clusters can be looked upon as the building blocks of starburst
galaxies, and should thus trace the IMF for rather extreme
environmental conditions, too. However, the IMF in starburst clusters
is not necessarily biased towards massive stars.

The first super star clusters with dynamical mass estimates were
NGC~1569A and NGC~1705-1 \cite{ho96}. With a mass of $1.1 \times 10^6$
$M_\odot$ (NGC~1569A) and $2.7 \times 10^5$ $M_\odot$ (NGC~1705-1),
these clusters are both gravitationally bound and could evolve into
globular clusters \cite{ho96}. However, recent evolutionary synthesis
modelling of the mass-to-light ratio \cite{sternberg98} revealed that
the IMF must be very different in the two regions. While the IMF in
NGC~1596A is steep (slope $\approx$ -1.5) and extends to below 0.1
$M_\odot$, the IMF in NGC~1705-1 is either flat (slope $>$ -1) or
truncated at a lower mass limit of 1 -- 3 $M_\odot$.

A similar conclusion has been found for the Antennae
\cite{mengel01}. The new generation of infrared spectrometers has
allowed the dynamical mass determination of young embedded super star
clusters for the first time. The authors \cite{mengel01} have compared
each mass measurement with the results with the stellar synthesis
model Starburst99 \cite{leitherer99}, that fit best the observed
NIR-luminosity, Br$\gamma$, and the CO and Ca absorption features. The
measurements in two of the observed clusters are consistent with a
Salpeter IMF extending to subsolar masses. However, in their cluster
\#2 \cite{whitmore99}, such a power law IMF with a slope of -1.35
would include all the dynamical mass ($\approx 1.6 \times 10^6$
$M_\odot$) in stars more massive than 1 $M_\odot$. Therefore no mass
is left for subsolar mass stars. This cluster seem to have formed
proportionally more massive stars.

In conclusion, some of the measured mass-to-light ratios in super star
clusters support the theoretical arguments for a top heavy IMF in
starbursts, and thus systematic variations in the IMF, but the same
kind of measurements also point to random variations in the IMF from
cluster to cluster, even those in similar environments.

\subsection{Extragalactic Star Formation Complexes}

Unfortunately there are only a few dynamical mass measurements for
star clusters. Statistics on IMF variations for a larger number of
extragalactic star forming regions has to rely on integrated
photometric and spectroscopic properties.

A recent example for such a statistical approach is \cite{sakhibov00},
who have included 105 extragalactic star formation complexes for which
UBVR photometry, Lyman continuum flux, metallicity and extinction
measurements were available. These properties trace essentially the
high mass stellar content, and the results are restricted to the IMF
for stars more massive than 10 $M_\odot$. The models include the slope
of the IMF, the maximum stellar mass and the age of the cluster. The
star formation history is assumed either as a delta burst or as
continuous. The average slope of the IMF was found to be -1.42.  The
standard deviation in the measured distribution is 0.91, and the
authors estimate the standard error of their method to be
0.51. Therefore the measured distribution of slopes is meaningful
\cite{sakhibov00}. Because the total number of analysed star formation
complexes is rather large, the formal probability of measuring such a
large standard deviation is basically negligible ($< 10^{-10}$), if
one assumes a universal IMF.

Another investigation \cite{mashesse99} has concentrated on 17 star
formation regions in a sample of blue compact and irregular galaxies,
using the integrated properties from UV, optical, FIR and radio
measurements. This analysis uses the equivalent widths of H$\beta$, Si
IV, and Ca IV to constrain the relative number of O3-O8 stars to B0-B3
stars, and thus the slope of the high mass IMF for stars more massive
than 10 $M_\odot$.  The evolutionary tracks of starburst regions with
three modelled IMF slopes of 0.0, -1.35, -2.0 are clearly separated in
a W(H$\beta$) over W(Si IV)/W(C IV) diagram. The derived slope for the
starburst regions varies between 0 (3 regions), -1.35 (8 regions) and
-2.0 (5 regions). Despite the variation in the derived slopes of the
IMF, the authors interpret their result as in supporting a universal
IMF, specifically because they see no trend with metallicity
\cite{mashesse99}. However, the scatter in the diagram is much larger
than the average error bars, and I would interpret the same diagram as
evidence for a varying IMF. In addition, the authors also report that
objects showing flatter IMFs are always small compact star forming
regions, that four of the galaxies with optical continuum dominated by
a previous generation of stars have IMF slopes close to -2, and that a
previous burst of star formation could have hampered the formation of
lower mass stars.

\section{IMF Variations in the Galaxy}

A direct measurement of the IMF from star counts is still restricted
to our Galaxy and the most nearby galaxies. While we can only
compare the relative number of high- to low-mass stars or constrain the
slope of the high-mass IMF in the early Universe and in extragalactic
starburst regions, we are able to trace the IMF down to substellar
masses in nearby star forming regions. Figure \ref{figure} shows a
comparison of several IMFs measured in the Galaxy. We find variations
in the slope and/or the shape of the IMF for all mass ranges.

\subsection{The Most Massive Star Forming Regions}

The most massive stars found in the Galaxy have about 100
$M_\odot$. Most of these stars have been found in either very compact
star clusters or rather loose OB associations. The observations of
these regions have improved significantly in the last few years with
the HIPPARCOS mission, the Hubble Space Telescope (HST) and adaptive
optics assisted telescopes.

For example, HIPPARCOS has revealed 178 new members of the Upper
Scorpius OB2 association \cite{brown98}. Only 91 stars were previously
known in this association. Figure \ref{figure} shows the preliminary
IMF. With a slope of $\approx$ -1.9, this IMF resembles very well
Scalo's \cite{scalo86} field star IMF. I have compared this high-mass
IMF with the IMF in NGC~3603, the most massive visible HII region in
our Galaxy \cite{eisenhauer98}. Adaptive optics assisted observations
have revealed more than 800 stars in the central parsec of this
cluster. Although the IMF shows no turnover or truncation down to $<
1$ $M_\odot$, its slope is only $\approx$ -0.7 for stars with masses
between 3 and 30 $M_\odot$. A very similar result was found with the
help of HST in the Arches cluster near the Galactic centre
\cite{figer99}. The slope of the IMF for stars more massive than 10
$M_\odot$ is $\approx$ -0.65, much shallower than the reference field
star IMF, too. The Arches cluster and NGC~3603 are two of the most
massive young clusters in the Galaxy, and each cluster has a few 1000
$M_\odot$ in O-stars. Therefore the difference between the field star
IMF and the IMF in NGC~3603 and the Arches cluster can not be
explained by small number statistics. Such an argument is often used
to invalidate the evidence for IMF variations from star counts
\cite{kroupa00}. The finding in these two clusters supports the
hypothesis that the IMF in starburst regions is biased towards massive
stars. However, the deficit of low-mass stars is caused by a shallow
IMF, and does not result from a truncated IMF, which has been
indicated in M~82 \cite{rieke93}.

\subsection{Associations and Open Clusters}

The discussion about a universal IMF was revived by Scalo
\cite{scalo98} in 1997, at a time, when the majority of the
participants of a conference on the stellar initial mass function was
arguing for a universal IMF \cite{gilmore98}. He made his case for IMF
variations basically from a statistical interpretation of the measured
slopes of the IMF in Galactic and Magellanic cloud associations and
open clusters. Scalo plotted the slope of the IMF of 61 clusters
against the average logarithmic stellar mass considered in the
measurements. Although the typical error in the slopes is 0.1 -- 0.4,
the spread in the diagram above $\approx 1$ $M_\odot$ is so large that
the author saw no basis for adopting some average value.

A serious argument against such evidence for IMF variations is that
the results for different clusters have been obtained from different
authors using different techniques. Therefore the IMF variations may
have been mimicked by systematic errors, which have not been addressed
carefully enough. I have thus repeated a similar $\chi^2$ analysis for
this conference to estimate how large the systematic errors would have
to be, if the IMF were to be universal. My compilation includes the
slopes of the IMFs for 51 clusters with intermediate- to high-mass
star formation from 7 publications. The publications have been
selected such that each paper includes more than 4 clusters, so that I
could carry out statistics for each author and technique
separately. Table \ref{table} summerises the IMF properties of the
different publications. Clusters with suspicious IMF measurements (for
example NGC~436 in \cite{phelps93}, which has probably undergone
dynamical evolution), and results without error estimates (for example
NGC~7235 in \cite{massey95a}) or with a different mass range (for
example NGC~376 and N~24 in \cite{hill94}) have not been included in
the statistics. For the ease of interpretation, I have converted the
reduced $\chi^2$ values into a significance level \cite{iso79}, which
is the probability of measuring such a high or higher $\chi^2$, if the
IMF would have a universal slope equal to the weighted mean slope in
this subsample. Even at a significance level of only 5 \%, we have to
reject the hypothesis of a universal IMF in 5 of 7 cases. The
statistics for the whole sample is even worse. The standard deviation
in the slope distribution is 1.9 times the mean error in the
measurements of the slopes. The likelihood for such a large discrepancy
is basically negligible ($< 10^{-10}$) for a universal IMF.  If the
IMF is indeed universal, the authors must have underestimated their
errors by a factor of two. The systematic errors would have to be even
larger if we include the field stars of the Magellanic clouds, which
show a particularly steep IMF with a slope of $\approx$ -3 to -4
\cite{massey95b}.

\begin{table}
\caption{IMF variations in 57 clusters with intermediate- to high-mass
star formation}
\begin{center}
\setlength\tabcolsep{5pt}
\begin{tabular}{llllllll}
\hline\noalign{\smallskip} 
Ref. & Number & Weighted & Standard & Mean error & Reduced & Significance \\
& of & mean & deviation & in slope & $\chi^2$ & \\
& clusters & slope & in slopes &  measurements & \\
\noalign{\smallskip} \hline \noalign{\smallskip} 
\cite{subramaniam99} & 4 & 1.43 & 0.34 & 0.15 & 5.15 & 0.15 \% \\
\cite{sagar98} & 5 & 1.27 & 0.80 & 0.33 & 14.82 & $\approx 0$ $^{\rm a}$ \\
\cite{oey96} & 6 & 1.22 & 0.31 & 0.30 & 2.92 & 1.2 \% \\
\cite{massey95a} & 12 & 1.12 & 0.44 & 0.34 & 1.78 & 5.3 \% \\
\cite{massey95b} & 5 & 1.27 & 0.18 & 0.16 & 1.64 & 16.2 \% \\
\cite{hill94} & 12 & 1.92 & 0.54 & 0.34 & 3.22 & 0.021 \% \\
\cite{phelps93} & 7 & 1.42 & 0.26 & 0.15 & 18.08 & $\approx 0$ $^{\rm a}$ \\
\noalign{\smallskip} \hline \noalign{\smallskip} 
all & 51 & 1.52 & 0.53 & 0.28 & 6.35 & $\approx 0$ $^{\rm a}$ \\
\end{tabular}
\end{center}
\label{table}
\footnotesize
$^{\rm a}$ The significance level is $< 10^{-10}$.
\end{table}

A very intriguing example with very different IMFs at intermediate
stellar masses are the Pleiades and the Orion nebular cluster. These
clusters have been the target of some of the most careful recent
measurements of the IMF \cite{meusinger96,hillenbrand97}. The IMFs of
these two clusters are also displayed in figure \ref{figure}. In
contrast to the IMF of the Pleiades, which follows the field star IMF,
the IMF in Orion shows a regime of "missing" stellar masses between
0.5 -- 1.25 $M_\odot$. Other examples are NGC~6231 and NGC~2264. The
IMFs in these clusters have been measured in both cases with the same
technique --- UBVRI and H$\alpha$ photometry --- and by the same
authors \cite{sung97,sung98}, therefore minimising systematic
variations.  Nevertheless, the IMF in NGC~2264 rises continously down
to below 0.6 $M_\odot$, whereas the IMF decreases abruptly below 2.5
$M_\odot$ in NGC~6231.

\subsection{Low-Mass Stars in Young Clusters, Globular Clusters and the
Field}

If we assume a Salpeter \cite{salpeter55} power-law IMF with a slope
of -1.35 extending down to a lower mass limit of 0.1 $M_\odot$, 96 \%
of all stars would have a mass smaller than 1 $M_\odot$, and 55 \% of
the stellar mass would be included in stars less massive than the sun.
However, these low-mass stars are very faint, and many such stars have
not been detected before the large ground based infrared surveys and
HST. These observations have also revealed IMF variations in many
regions.

To illustrate these variations, I will again compare different
regions with the field star population. Figure \ref{figure} shows a
recent measurement \cite{reid99} of the field star IMF for low-mass
stars. It was derived from the Deep Near-Infrared Survey (DENIS) and
the 2 Micron All-Sky Survey (2MASS). In the $0.1 - 1$ $M_\odot$ mass
range, this IMF can be represented by a power-law mass function with a
slope of $\approx 0.1$. The statistical uncertainties are $\approx
0.13$. A similar IMF has been observed in the young star forming
regions IC~348, $\rho$~Ophiuchi, and the Trapezium
\cite{luhman98,luhman99,luhman00a}. Their IMF is flat or slowly rising
--- slope
$\mathrel{\hbox{\rlap{\hbox{\lower4pt\hbox{$\sim$}}}\hbox{$>$}}}$ 0
--- from the brown dwarf regime to $0.6 - 1$ $M_\odot$, where it rolls
over to a power-law with a slope of $\approx -1.7$. It is important to
note that the IMF is not log-normal \cite{luhman00a}.

In contrast, recent HST observations indicate that the IMF of globular
clusters can be described by a log-normal function with a peak near
$0.3$ $M_\odot$ \cite{paresce00}. Figure \ref{figure} shows the best fit
for a sample of 12 globular clusters. The authors \cite{paresce00}
explicitly exclude a single power-law IMF in the $0.1 - 0.6$ $M_\odot$
range. Therefore the IMF of globular clusters seems to be
fundamentally different from the IMF in young star forming regions and
the field.

A meaningful interpretation of the IMF measurements for masses below
the hydrogen burning limit is increasingly more difficult, because the
available data are sparse. However, observations of young star forming
regions indicate IMF variations even in the mass regime of brown
dwarfs. The Taurus star forming region, for example, is significantly
deficient in objects below 0.1 $M_\odot$ compared to the Trapezium
\cite{luhman00b}. If the IMF of both objects is normalised by the
number of stars between 0.1 - 1 $M_\odot$, then $\approx 13$ brown
dwarfs with masses $> 0.02$ $M_\odot$ are found in the Trapezium, but
only one in Taurus.

\section{Conclusions}

Variations of the stellar initial mass function have been reported for
all masses and in a large variety of stellar populations. We find
evidence for IMF variations in the early Universe, in starburst
galaxies and extragalactic star forming regions, in Galactic star
clusters and associations, and in the field star population. Neither
the shape nor the slope and the characteristic mass of the IMF seem
to be excluded from these variations. There is some indication that
the IMF is systematically biased towards more massive stars in the
early Universe and in starbursts. However, I see no obvious systematic
trend in those regions where the IMF could be constructed from direct
star counts. If the IMF is indeed universal, it will be very difficult
to prove this postulation empirically, because every individual
measurement of a different IMF has to be invalidated. Only a
consistent theory of star formation, with clear and testable
predictions, will finally convince in the scientific world.

\end{document}